\documentclass[prl,twocolumn,,amsmath,amssymb,floatfix,superscriptaddress]{revtex4}
\usepackage{graphicx}
\usepackage{times}

\begin{document}
\title{Invisibility's Flicker: Detecting Thermal Cloaks via Transient Effects}

\author{Sophia R. Sklan}
\affiliation{Department of Physics, Massachusetts Institute of Technology, Cambridge, Massachusetts 02139, USA}
\affiliation{Department of Mechanical Engineering, University of California, Berkeley, California 94720, USA}
\author{Xue Bai}
\affiliation{Department of Electrical and Computer Engineering, National University of Singapore, 4 Engineering Drive 3, Singapore 117583, Republic of Singapore}
\affiliation{Department of Physics and Centre for Computational Science and Engineering, National University of Singapore, Singapore 117546, Republic of Singapore}
\affiliation{NUS Graduate School for Integrative Sciences and Engineering, National University of Singapore, Kent Ridge 119620, Republic of Singapore}
\author{Baowen Li$^*$}
\affiliation{Department of Physics and Centre for Computational Science and Engineering, National University of Singapore, Singapore 117546, Republic of Singapore}
\affiliation{NUS Graduate School for Integrative Sciences and Engineering, National University of Singapore, Kent Ridge 119620, Republic of Singapore}
\affiliation{Center for Phononics and Thermal Energy Science, School of Physics Science and Engineering, Tongji University, 200092, Shanghai, China. \\ *e-mail: phylibw@nus.edu.sg}
\affiliation{Department of Mechanical Engineering, University of California, Berkeley, California 94720, USA}
\author{Xiang Zhang}
\affiliation{Department of Mechanical Engineering, University of California, Berkeley, California 94720, USA}
\affiliation{NSF Nanoscale Science and Engineering Centre, 3112 Etcheverry Hall, University of California, Berkeley, California 94720, USA}
\affiliation{Materials Sciences Division, Lawrence Berkeley National Laboratory, 1 Cyclotron Road, Berkeley, California 94720, USA}

%%% \date{\today}

\begin{abstract}
Recent research on the development of a thermal cloak has concentrated on engineering an inhomogeneous thermal conductivity and homogeneous volumetric heat capacity.
While the perfect cloak of inhomogeneous $\kappa$ and $\rho c_p$ is known to be exact (no signals scattering or penetrating to the cloak's interior), no such analysis has been considered for this case.
Using analytic, computational, and experimental techniques, we demonstrate that these approximate cloaks are detectable.
Although they work as perfect cloaks in the steady-state, their transient (time-dependent) response is imperfect and a detectable amount of heat is scattered.
This is sufficient to determine the presence of a cloak and any heat source it contains, but the material composition hidden within the cloak is not detectable in practice.
\end{abstract}

\maketitle

The ability to render an object invisible has been a goal since the days of mythology and the Ring of Gyges.
It is only recently that invisibility became a plausible subject of inquiry thanks to theoretical advances in electromagnetism \cite{TO1,TO2,EMC1,EMC2,EMC3,EMC4}.
 Such cloaks fulfilled the two basic elements of invisibility: anything hidden inside was isolated as if hidden by a perfect insulator and the perfect insulator had no scattering.
This first requirement typically entailed singular, anisotropic materials, while the second required inhomogeneity.
These extreme material requirements turned attention to reduced cloaks which merely approximate perfect cloaking \cite{RC1,RC2,RC3,RC4}
 or conditions where these constraints are relaxed \cite{CC1,CC2,CC3,CC4,CC5,CC6,CC7}.
 This in turn led to the study the detectability of these cloaks \cite{PC1,PC2,PC3,PC4,PC5}.
 Concurrently, cloaking was extended to other classes of electromagnetic phenomena \cite{VC1,VC2,EMCR1,EMCR2,EC1,MC1,MC2},
wave equations \cite{WC1,WC2,WC3,WC4,WC5,WC6,WC7,WC8}
, and diffusion equations \cite{QMC1,QMC2,QMC3,TC1,TC2,TC3,TC4,TC5,TC6,TC7,BC,CDC1,CDC2,EMD}
(\cite{NEMR,Rev2} provide a review of these last categories).

The diffusive cloaks found greatest success with the heat equation:
\begin{equation}\label{eq:HE}
\rho c_p \partial_t T = \nabla\cdot\left(\kappa\nabla T\right)
\end{equation}
where $\rho$ is the density, $c_p$ the specific heat capacity, $T$ temperature, and $\kappa$ the thermal conductivity, so we shall confine our attention to thermal cloaks and then generalize our results to other diffusion effects.
Because the steady-state temperature is independent of the volumetric heat capacity $\rho c_p$, cloaking has focused on engineering $\kappa$ with $\rho c_p$ constant.

\begin{figure}%[th!]
\begin{center}
%\vspace*{5cm}
\includegraphics[scale=0.6]{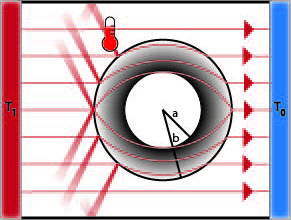}
\caption{\label{fig:model} Simple model for detecting a cloak. Cloaked region of radius $a$ inside a cloak of radius $b$ is protected from outside searchers. To detect the cloak, it is surrounded by two heat baths (red and blue rectangles) in a thermally isolated domain. Heat (red lines) flows through the cloak and emerges without distortion. However, for an imperfect cloak heat is also scattered. The scattered heat diffuses, but a thermometer placed near the cloak can detect it.
}
\end{center}
\end{figure}

In this paper, we show a homogeneous $\rho c_p$ results in a detectable, transient signal. 
Under changing boundary conditions, a thermal cloak will flicker and become visible, although it will help to obscure anything hidden inside it.
The implications of this imperfection can be seen by considering a faulty cloak with one observer hidden inside and another searching outside (see Fig. \ref{fig:model}).
The searcher can send out signals and detect the diffuse ``scattering'' that is reflected back.
Furthermore, they can search for signals emanating from the cloak's interior $-$ thereby determining the material composition or temperature distribution hidden inside.
Conversely, the observer hiding in the cloak can detect incoming signals to observe any searchers and eavesdrop on outside.
Moreover, by sending out their own signals and detecting them, they can confirm that the cloak is present and functioning.

%\section{Results}
We begin by considering the analytic solution to eq. \ref{eq:HE} for the cylindrical perfect cloak (PC) (assuming no $z$ dependence).
For a homogeneous medium ($\kappa=\kappa_0, \rho c_p = \rho_0 c_{p0}$), source-free medium the solution can be expressed as a linear combination of the fundamental solutions
\begin{equation}
T_l(r,\theta,\omega) = R_l(\sqrt{i}k_0r)e^{il\theta+i\omega t}
\end{equation}
where $l$ the rotational symmetry, $R_l$ is a modified Bessel function of the first or second kind ($I_l$ and $K_l$ respectively), $k_0=\sqrt{\omega\rho_0 c_{p0}/\kappa_0}$, and $\omega$ is a frequency $>$0 (the steady state of $\omega=0$ is discussed in the supplement).
A PC of interior radius $a$ and exterior radius $b$ (Fig. \ref{fig:model}) is constructed from
\begin{gather}
\kappa_r=\kappa_0\frac{r-a}{r},\ \ \ \ \kappa_\theta=\kappa_0\frac{r}{r-a}, \nonumber	\\
\rho c_p=\rho_0c_{p0}(\frac{b}{b-a})^2\frac{r-a}{r}
   \label{eq:cloak}
   \end{gather}
the solution becomes
\begin{equation}
	T_l^{(PC)}(r,\theta,\omega) = R_l(\sqrt{i}k_C[r-a])e^{il\theta+i\omega t}
	\label{eq:PCsol}
\end{equation}
where $k_C/k_0=b/(b-a)$.
Whereas, for a steady-state cloak (SSC) $\kappa$ is the same as eq. \ref{eq:cloak} but $\rho c_p = (b/(b-a)) \eta \rho_0 c_{p0}$ ($\eta$ a mismatch parameter, $\eta=1$ is eq. \ref{eq:cloak} evaluated at $r=b$), the lowest order perturbation is
\begin{eqnarray}
	T_l^{(SSC)}(r,\theta,\omega) &=& R_l(\sqrt{i\eta}k_S[r-a])e^{il\theta+i\omega t} \nonumber \\
		&+& \sqrt{i} \lambda \mathcal{F}\left[R_l(\sqrt{i\eta}k_S[r-a])\right]e^{il\theta+i\omega t}
	\label{eq:SSCsol}
\end{eqnarray}
where $k_S/k_0=\sqrt{b/(b-a)}$, $\lambda=\sqrt{\eta} k_S a$ and $\mathcal{F}[R]$ is given in the supplement (along with the analytic solution).
Crucially, $\lambda$ determines the strength of the perturbation, meaning the effectiveness of an impedance matched SSC ($\eta=1$) is proportional to the size of the cloaked region over the diffusion length.
For reference, the various cloaks considered in this paper are summarized in Table \ref{tbl:cloaks}

\begin{table}
\begin{tabular}{|c|c|c|c|}
\hline 
Cloak & $\kappa_r/\kappa_0$ & $\kappa_\theta/\kappa_0$ & $\rho c_p/\rho_0 c_{p0}$\tabularnewline
\hline
\hline 
PC & $(r-a)/r$ & $r/(r-a)$ & $[b/(b-a)]^2 (r-a)/r$\tabularnewline
\hline 
SSC (M) & $(r-a)/r$ & $r/(r-a)$ & $b/(b-a)$\tabularnewline
\hline 
SSC(Mis) & $(r-a)/r$ & $r/(r-a)$ & $[b/(b-a)]^2$\tabularnewline
\hline 
BC $\{r\in (a,r_1)\}$	& $\kappa_1/\kappa_0$ & $\kappa_1/\kappa_0 $ & $\rho_1 c_{p1}/\rho_0 c_{p0} $\tabularnewline
%\hline 
BC $\{r\in (r_1,b)\}$	& $\kappa_2/\kappa_0$ & $\kappa_2/\kappa_0 $ & $\rho_2 c_{p2}/\rho_0 c_{p0}$\tabularnewline
\hline 

\end{tabular}
\caption{\label{tbl:cloaks} Equations for cloaks considered in this paper. Perfect cloak (PC), impedance matched steady-state cloak (SSC (M)), impedance mismatched SSC (SSC (Mis)) ($\eta\equiv b/(b-a)$), and bilayer cloak (BC). Inner layer of the cloak has radius $a$, outer layer radius $b$, as in Fig. \ref{fig:model}.}
\end{table}

Since the solution to eq. \ref{eq:SSCsol} is not a tabulated function, we use COMSOL multiphysics to solve eq. \ref{eq:HE} directly for the SSC.
Following the most common test of a cloak, we model the cloak in a rectangular region where one pair of ends are held at fixed temperature and the other pair admit no heat (Fig. \ref{fig:model}).
Given the linearity of eq. \ref{eq:HE}, one boundary is set to 0 (as is the initial $T$) and the other to 1 ($\Delta T\equiv 1$).
It is helpful to use the natural units of $L$ (the separation of the heat sources) and the diffusion time $\tau_D=L^2 \rho_0 c_{p0}/\kappa_0$ (all parameters  values are in the supplement).
Fig. \ref{fig:SSCdeviation} is the result of these calculations.
Each column is a snapshot at a different time.
The first row is the homogeneous background that would be observed if there was no cloak, the second is the solution to SSC (with $\eta=b/(b-a)$ to increase contrast), and the third is the difference $\delta T=T^{(SSC)}(\vec{r},t)-T^{(H)}(\vec{r},t)$.
This deviation $\delta T$ is what must be detected to reveal a cloak.
Initially $\delta T$ is small and mostly localized to where the cloak has been heated (see Fig. \ref{fig:SSCdeviation}g).
Later (Fig. \ref{fig:SSCdeviation}h) $\delta T$ grows and is clearly observed outside the cloak.
Finally in the steady state (Fig. \ref{fig:SSCdeviation}i) invisibility is restored, as expected for a SSC ($\delta T\ne0$ confined to within the cloak).

\begin{figure}%[h!]
\begin{center}
\includegraphics[scale=0.1]{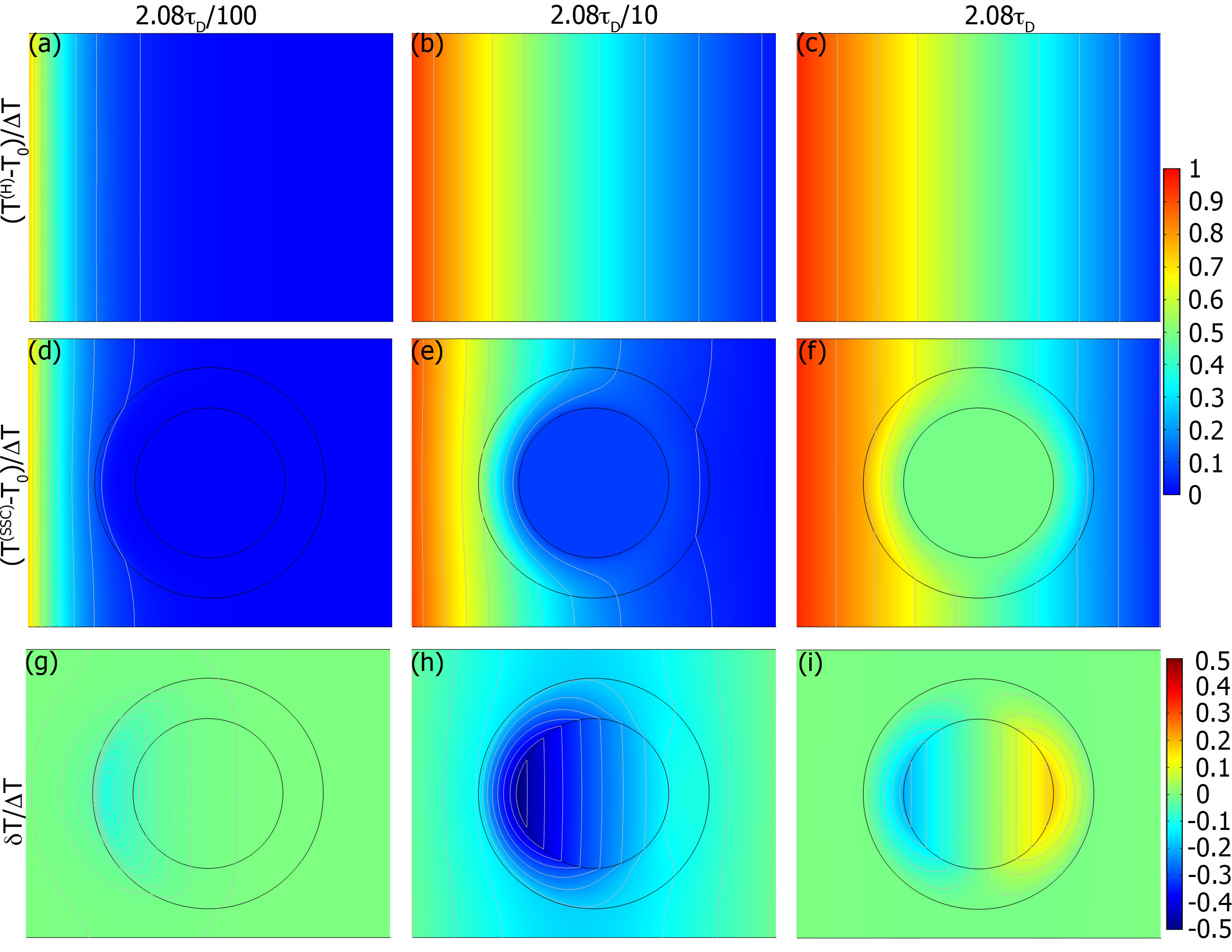}
\caption{\label{fig:SSCdeviation} Simulated temperature snapshots for mismatched SSC ($\eta = b/(b-a)$). Rows correspond to 2.08$\tau_D$/100, 2.08$\tau_D$/10, and 2.08$\tau_D$ respectively. Columns correspond to the homogeneous case (no cloak), SSC, and $T^{(SSC)}-T^{(H)}$. Black circles denote the location of the cloak (for reference in the homogeneous case), colored domains are isotherms, and grey lines are constant separation isotherms.
}
\end{center}
\end{figure}

To clarify the time-dependence of $\delta T$ we select several points outside the cloak and compare $\delta T$ for SSC with $\eta=1$ (i.e. impedance matched, cloak has the same properties as PC at $r=b$), $\eta = b/(b-a) > 1$ (impedance mismatch but $\sqrt{\eta}k_S=k_C$), and the PC in Fig. \ref{fig:time}.
As we prove in the supplement, $\delta T=0$ outside the cloak for the PC, so the non-zero $\delta T$ must be a numerical artifact of discretizing $\kappa$ and thereby removing $\kappa_\theta\to\infty$ (this is corroborated by the penetration of heat into the cloaked region).
However, for both SSC models $\delta T^{(SSC)}>\delta T^{(PC)}$ outside the cloak.
This is the effect a homogeneous $\rho c_p$.
Note that the position dependence of $\delta T$ is approximately just a scaling factor so that they peaks all nearly coincide instead of being separated by a propagation time.
Both this and the time dependence (linear growth for small $t$, exponential decay for large) are derived in the supplement, where we show that this separable space dependence implies that $\delta T$ is dominated by a small number of Fourier-modes.

\begin{figure}%[!h] 
\begin{center}
%\cleardoublepage
%\vspace*{7cm}
\includegraphics[scale=0.4]{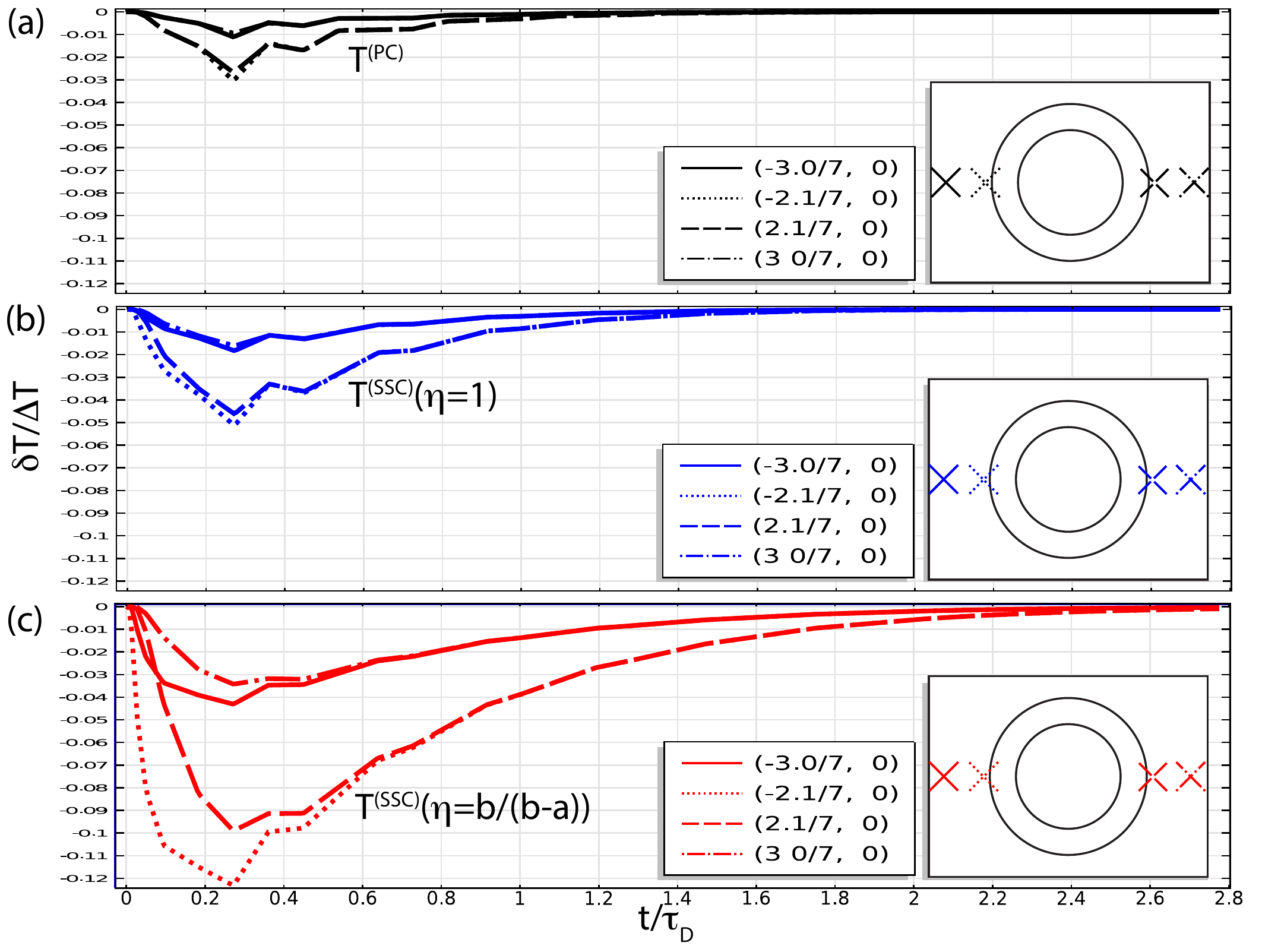}
%\vspace*{-1cm}
\caption{\label{fig:time}Temperature deviation $\delta T/\Delta T$ for representative points outside the cloak as a function of time.  Black (a), blue (b), and red (c) curves correspond to the PC, impedance matched SSC, and impedance mismatched SSC. Line styles correspond to individual points, as shown in the inset. 
}
%\vspace{-3cm}
\cleardoublepage
\end{center}
\end{figure}

\begin{figure}%[p]
\begin{center}
%\cleardoublepage
%\vspace*{5cm}
%\newpage
%\pagebreak[4]
%\vspace*{-10cm}
%\pagebreak[4]
%\\
%\\*
%\\
%\vspace{5cm}
\includegraphics[scale=0.18]{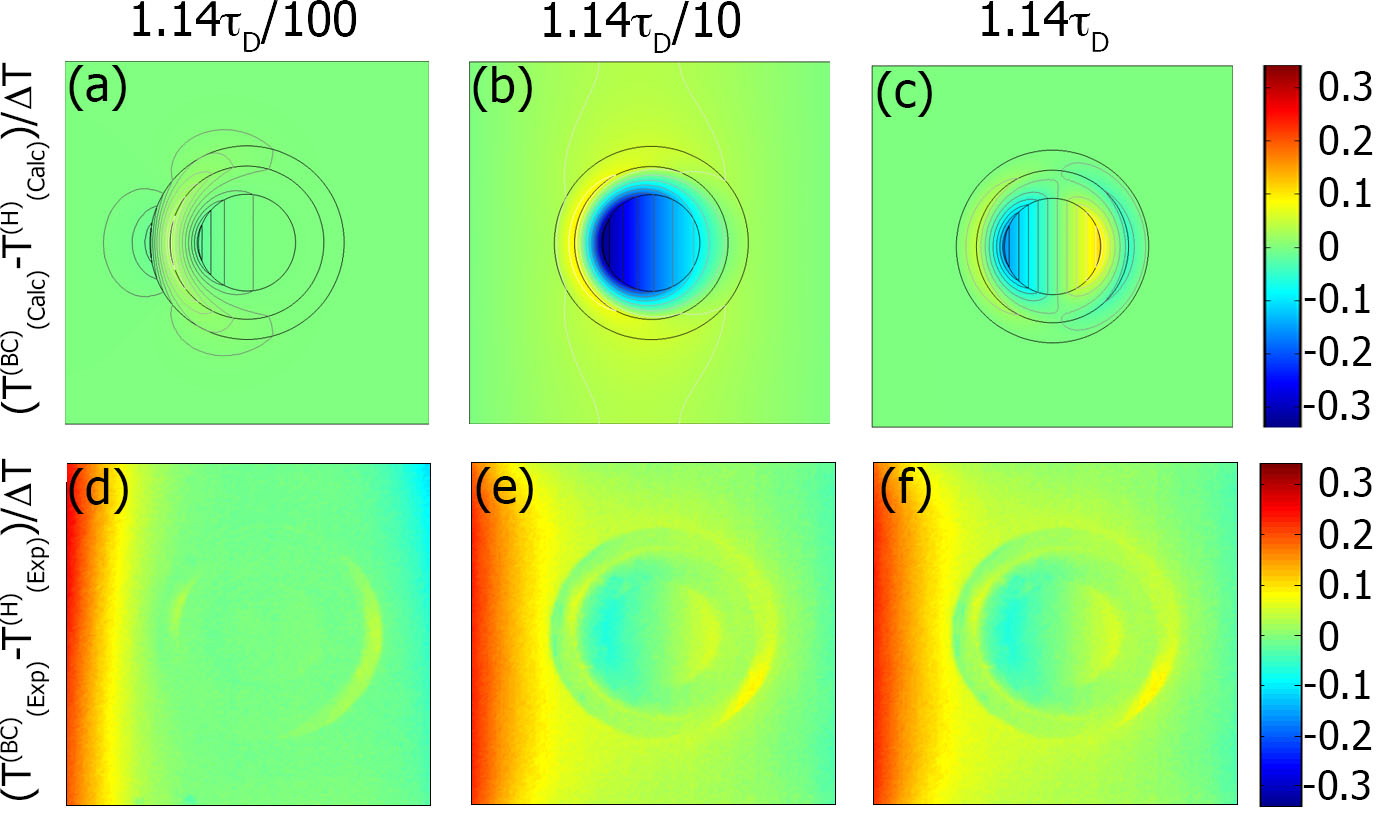}
\caption{\label{fig:BCC} Comparison of simulations and experimental for the BC. Rows correspond to $1.14\tau_D/100$, $1.14\tau_D/10$, and $1.14\tau_D$ respectively. Columns correspond to $\delta T$ for the simulation and experiment respectively.
}
%\cleardoublepage
\end{center}
\end{figure}

To compare our theoretical predictions for the SSC with its inhomogeneous $\kappa$ with the behavior of an experimentally realized SSC with discretized rings of constant $\kappa$ we consider the bilayer cloak (BC) \cite{BC} (simulations and further experimental data for the BC are in the supplement).
The BC is particularly interesting to consider as it is a SSC that was derived directly from Laplace's equation rather than a coordinate transformation.
In Fig. \ref{fig:BCC} we plot the normalized temperature deviation for the simulated BC and our experimental realization.
This shows a good agreement, with a slight discrepancy near the boundaries of the system.
This is due to a slight difference in the experimental temperature gradients applied to the BC and homogeneous cases.

Finally, we turn to the question of detecting objects hidden inside a cloak.
For the PC and the SSC $\hat{r}\cdot\kappa\nabla T=0$ at the boundary ($\kappa_r=0$), so there should be no heat transferred and therefore no discernable signal (although, as in \cite{PC1}, this is extremely sensitive to deviations of $\kappa_r$ from 0).
However, taking the BC and changing the material hidden inside will effect the temperature distribution.
An exterior temperature profile like those considered above must pass through the cloak twice (entering and exiting), so the cloaks ability to suppress detection is stronger here than in the case of hiding the cloak.
In particular simulations (see supplementary materials) indicate that a gradient of over 100K would be necessary for our detectors.
On the other hand, heat initially confined to the cloak would diffuse out and only pass through the cloak once.
As we see in the supplement, this is a weaker signal than the case of detecting the cloak, but is in principle observable.
Our simulations also indicate that the BC is no more effective at suppressing this signal than a thermal insulator of thickness equal to insulating layer in the cloak.
This suggests that realizable cloaks (i.e. those without a perfectly insulating inner boundary) are no better than conventional insulators for maintaining a temperature difference.

We have shown that a SSC can be detected by its transient response.
Because the distinction between a PC and a SSC is just $\rho c_p$, the ability to engineer the volumetric heat capacity is necessary to prevent the $\omega\ne0$ response from revealing the cloak.
However, the narrow range of $\rho c_p$ in currently available materials makes it extremely difficult to design this inhomogeneity (indeed, even efforts to construct a ``transient'' thermal cloak have assumed constant $\rho c_p$ \cite{TC3,TC5}).
This is particularly true for other classes of diffusion cloaks where the analog of $\rho c_p$ is necessarily constant everywhere \cite{CDC1,CDC2}.
It remains an open question, however, if a diffusive cloak (thermal or otherwise) could be designed to make its time-dependent response undetectable in practice even if the response exists in principle.

This material is based upon work supported by the National Science Foundation Graduate Research Fellowship under Grant No. 1122374. 

\section{Supplementary Materials}
\makeatletter 
\renewcommand{\thefigure}{S\@arabic\c@figure}
\makeatother

\subsection{Scattering Solution to the Heat Equation}

Given the heat equation with homogeneous materials
\begin{equation}
\rho c_p \partial_t T = \nabla\cdot\left(\kappa\nabla T\right)
\end{equation}
in polar coordinates we take the Fourier transform of time and use a separable solution $T(r,\theta,t)= R(r)e^{i l \theta}e^{i\omega t}$ giving
\begin{equation}
\frac{i\omega \rho_0 c_{p0}}{\kappa_0} R = \frac{1}{r}\frac{d}{dr}(rR^{\prime})-\frac{l^{2}}{r^{2}}R.
\end{equation}
This is the differential equation for a modified Bessel function ($I_l(z)$ or $K_l(z)$) of $z=\sqrt{\frac{i\omega \rho_0 c_{p0}}{\kappa_0}}r$ for $\omega\ne0$.
The time-dependent solution is therefore
\begin{equation}
	T_l^{(tr)}(r,\theta,\omega) = \left(a_l I_l(z)+b_l K_l(z)\right)e^{il\theta+i\omega t}
\end{equation}

For the steady state of $\omega=0$ the solutions become the solution to Laplace's equation
\begin{equation}
T^{(SS)}_l(r,\theta) =\left(A_l r^l + B_l r^{-l}\right)e^{il\theta} 
\end{equation}
for $l\ne0$ and
\begin{equation}
T^{(SS)}_0 = A_0+ B_0 \ln(r)
\end{equation}
for $l=0$.
The general solution is therefore $T(r,\theta,\omega) = \Sigma_{l=0}^\infty T^{(SS)}_l+T^{(tr)}_l$.

For a perfect cloak
\begin{gather}
\kappa_r=\kappa_0\frac{r-a}{r},\ \ \ \ \kappa_\theta=\kappa_0\frac{r}{r-a}, \nonumber	\\
\rho c_p=\rho_0c_{p0}(\frac{b}{b-a})^2\frac{r-a}{r}
\end{gather}
we can make the coordinate transformation
\begin{equation}
r^\prime  =\frac{b}{b-a}(r-a)
\end{equation}
to reduce the solution in the primed coordinates to the homogeneous case.

For a steady-state cloak ($\kappa$ as for the perfect cloak, $\rho c_p=\rho_0 c_{p0} (b/(b-a)) \eta$, i.e. evaluating $\rho c_p$ at $r=b$ when $\eta=1$) no transformation will reproduce a homogeneous solution. Using $x=\sqrt{i \omega \rho_0 c_{p0} \eta b/\kappa_0(b-a)}(r-a)$ and separation of variables we find
\begin{equation}
0=\partial_x(x\partial_x R)-\left[\frac{l^2}{x}+x+Ka\right]R
\end{equation}
where $K=\sqrt{i \omega \rho_0 c_{p0} \eta b/\kappa_0(b-a)}$.
This can be solved by the method of Frobenius $R_l(x)=\Sigma b_{nl}^\pm x^{n\pm l}$ with recurrence relation
\begin{equation}
b_{nl}^\pm=\frac{1}{n(n\pm2l)}(Kab_{n-1,l}^{\pm}+b_{n-2,l}^{\pm}).
\end{equation}
This relation is exact, but additional insight can be gained by expanding the solution by powers of $Ka$.
For even terms in the series this is
\begin{equation}
b_{2m,l}^{\pm(0)} = \frac{1}{2m(2m\pm2l)}b_{2m-2,l}^{\pm(0)}+O\left([Ka]^{2}\right)
\end{equation}
which is the same as series expansion for $I_l$ and $K_l$ respectively.
On the other hand, for odd terms it becomes
\begin{equation}
b_{2m+1,l}^{\pm(0)} = Ka\sum_{n=0}^{m}\frac{|2n-1|!!}{(2m+1)!!}\frac{(2n\pm 2l-1)!!}{(2m\pm2l+1)!!}b_{2n,l}^{\pm(0)}+O\left([Ka]^{3}\right)
\end{equation}
Because $b_{2m+1,l}^{\pm(0)}$ is completely determined by $b_{2n,l}^{\pm(0)}$ the odd terms are therefore a function of the modified Bessel functions.
Ergo, we term these components $\mathcal{F}[R_l(x)]$.
A similar derivation can be carried out for a spherical cloak where $l$ becomes half-integer instead of integer.

\subsection{Simulations of the SSC}
\subsubsection{Simulation Details}

We model a rectangular domain of dimensions $L=$70 mm by $L_\perp=$50 mm centered around a cloak of dimension $a=$13 mm, $b=$20 mm.
The background medium is $\kappa_0=71.4 \mathrm{W/m\cdot K}$, $\rho_0 = 2100 \mathrm{kg/m^3}$, and $c_{p0} = 1000\mathrm{J/kg\cdot K}$.
This gives a diffusivity of $D=\kappa_0/\rho_0c_{p0}= 3.4\cdot10^{-5}\mathrm{m^2/s}$ and diffusion timescale $\tau_D=L^2/D=144.12$s.
The initial temperature was 293.15K with thermal baths at 300K, and $T_0=$293.15K giving a $\Delta T$ of 6.85K.
After confirming that the simulations were invariant under a change of scale we use the natural units of $x/L, y/L, t/\tau_D,(T-T_0)/\Delta T$.

%\subsection{Simulation of the PC}

%To confirm that the simulations worked as expected, we plot the solution for the PC in Fig. \ref{fig:PC}.
%Each column is a snapshot at a different time.
%The first row is the homogeneous background that would be observed if there was no cloak, the second is the solution to PC, and the third is the difference $\delta T=T^{(PC)}(\vec{r},t)-T^{(H)}(\vec{r},t)$.
%The deviation $\delta T$ is seen to be non-zero but confined within the cloak.
%This is as expected, for the cloak must distort the temperature distribution around the hidden region while leaving the exterior domain unperturbed.
%\begin{figure}[h] \begin{center}
%\includegraphics[scale=0.25]{T_n-T0_3s.png}
%\includegraphics[scale=0.25]{T_n-T0_30s.png}
%\includegraphics[scale=0.25]{T_n-T0_300s.png} \\
%\includegraphics[scale=0.25]{T_e-T0_3s.png}
%\includegraphics[scale=0.25]{T_e-T0_30s.png}
%\includegraphics[scale=0.25]{T_e-T0_300s.png} \\
%\includegraphics[scale=0.25]{T_e-T_n_3s.png}
%\includegraphics[scale=0.25]{T_e-T_n_30s.png}
%\includegraphics[scale=0.25]{T_e-T_n_300s.png}
%\includegraphics[scale=0.25]{T_e-T_n.pdf}
%\caption{\label{fig:PC} Simulated temperature snapshots. Rows correspond to %2.08$\tau_D$/100, 2.08$\tau_D$/10, and 2.08$\tau_D$ respectively. Columns %correspond to the homogeneous case (no cloak), PC, and $T^{(PC)}-T^{(H)}$. Black %circles denote the location of the cloak (for reference in the homogeneous case), %colored domains are isotherms, and grey lines are constant separation isotherms.
%}
%\end{center}
%\end{figure}

\subsubsection{Space Dependence of the Deviation of the SSC}

\begin{figure}[h] \begin{center}
\includegraphics[scale=0.35]{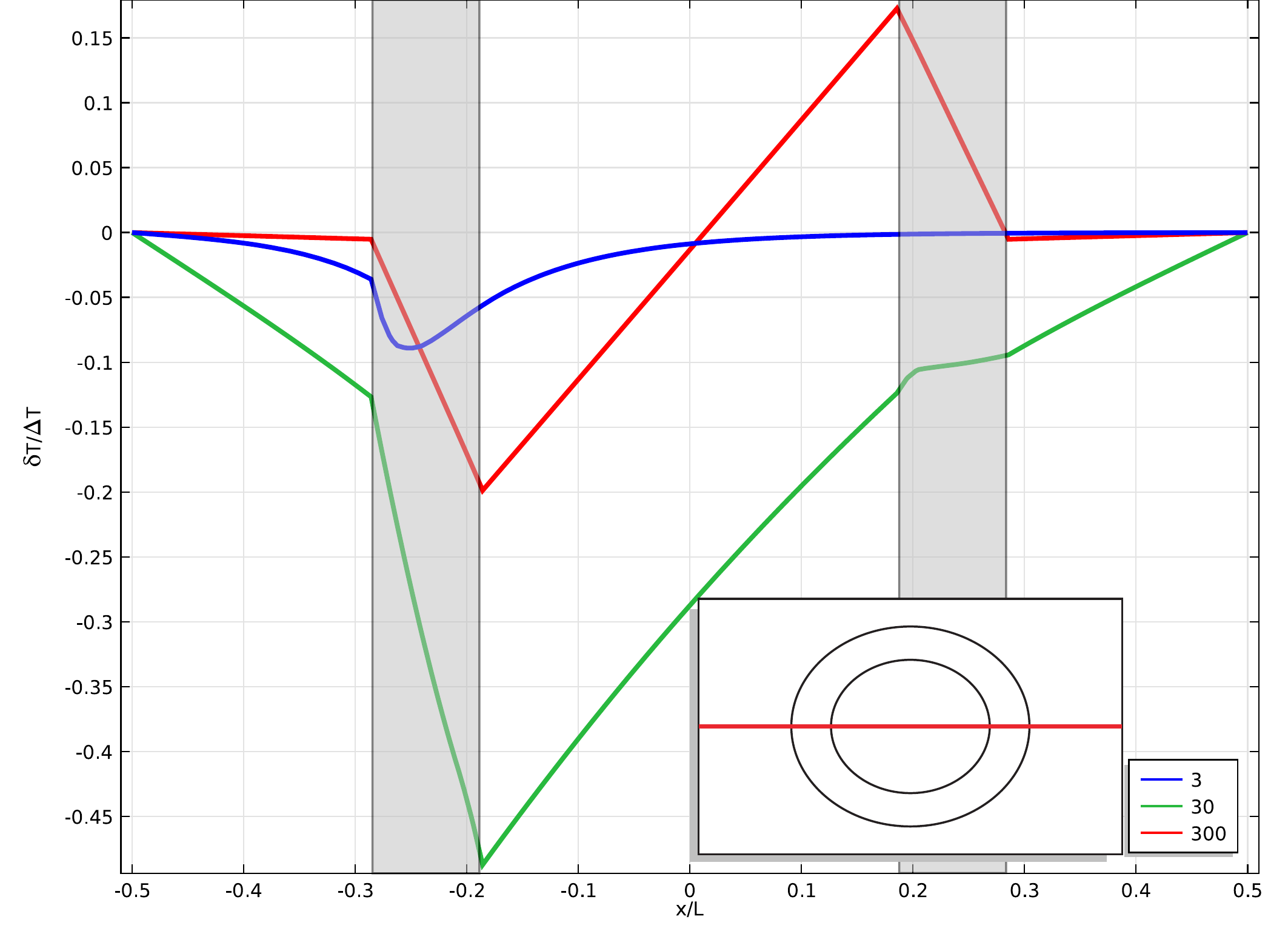}
\includegraphics[scale=0.35]{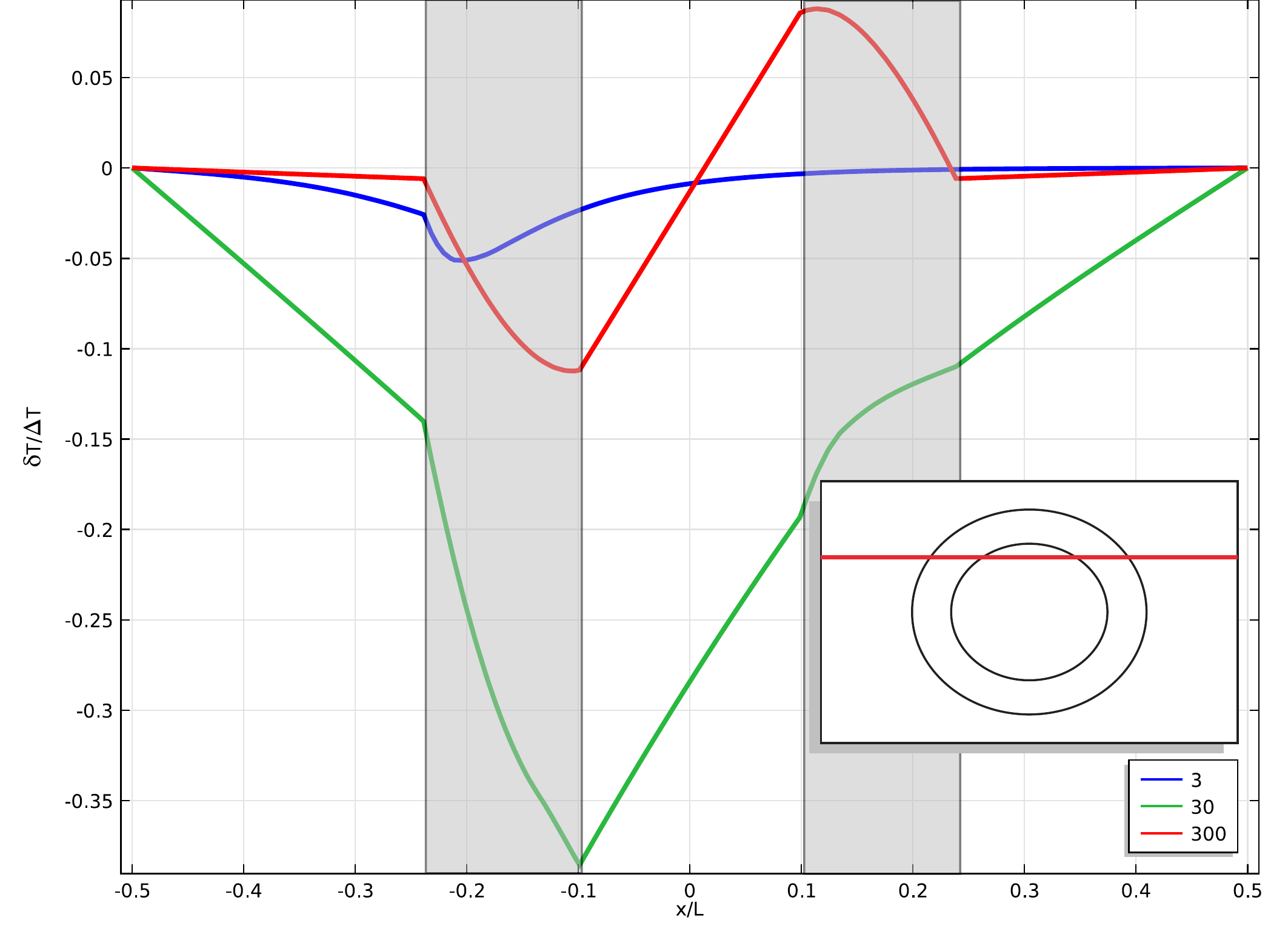}
\includegraphics[scale=0.35]{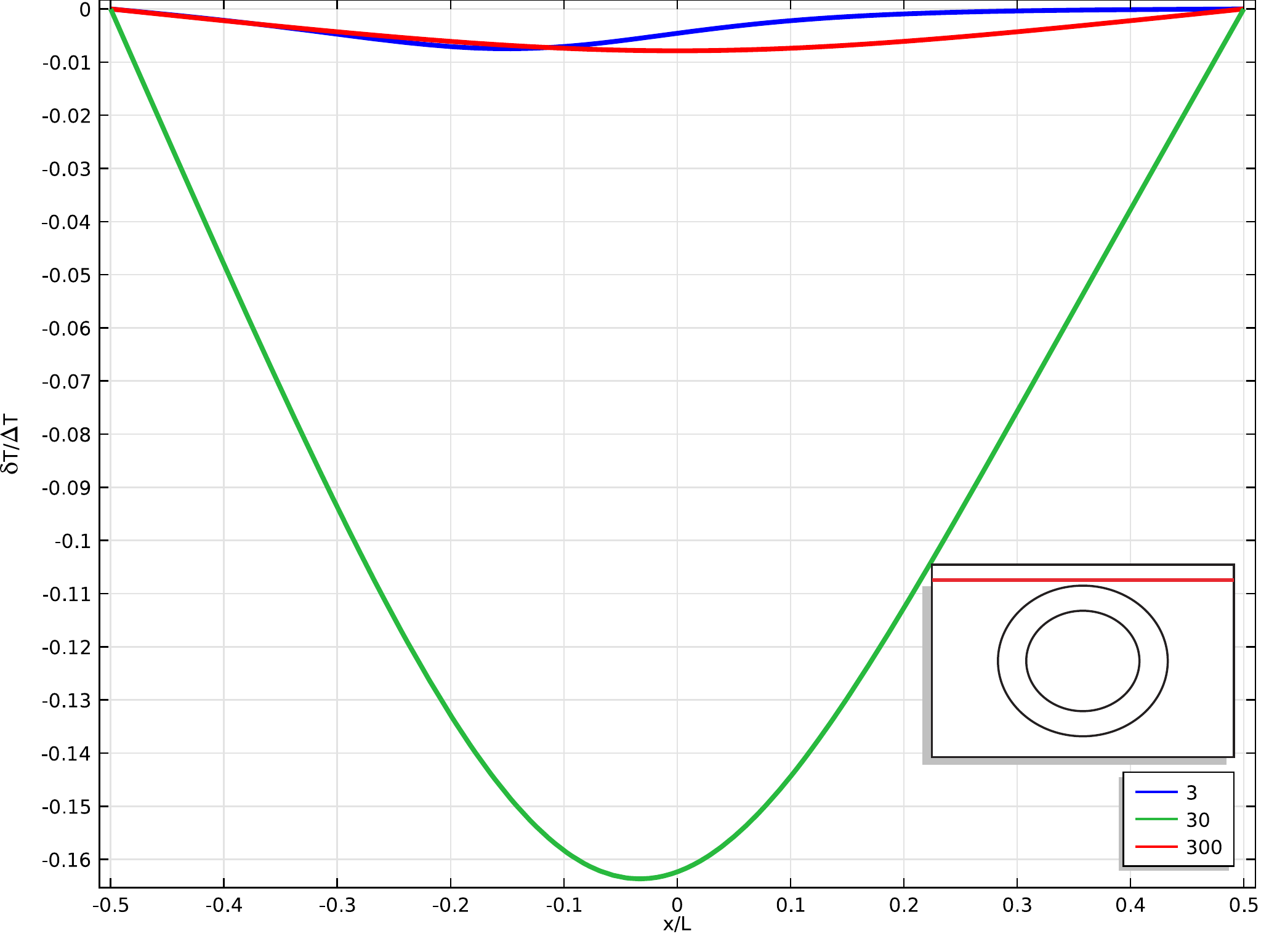} 
\caption{\label{fig:SSCspace} Space dependence of $\delta T$ of the SSC. Slices are along the middle of the cloak (plot (a), $y=0$), slightly offset from the center (plot (b)$y=1.1/7$), and outside the cloak (plot (c) $y=2.2/7$). The blue, green, and red curves are at 3s ($2.08\tau_D/100$), 30s ($2.08\tau_D/10$), and 300s ($2.08\tau_D$) respectively.
}
\end{center}
\end{figure}

In Fig \ref{fig:SSCspace} we take several slices of $\delta T$ along $y=$constant for $t=$ 2.08$\tau_D$/100, 2.08$\tau_D$/10, and 2.08$\tau_D$ (or 3s, 30s, and 300s)  (blue, green, and red respectively) to observe the spatial dependence more precisely.
Slices are centered, offset, and outside the cloak.
Initially the perturbation is well confined to the portion of the cloak that has been reached by the applied heat current.
As time passes and heat has spread relatively far into the domain the $\delta T$ grows and spreads throughout the domain.
As the system approaches steady state, $\delta T$ falls.
The linear dependence inside the cloak for steady state implies that $T^{(SSC)}$ inside this domain is essentially constant.
Outside the cloak $\delta T$ is effectively a sine curve.
This is clearest for the slice outside the cloak (after the initial curve, which contains higher that decay faster than the fundamental mode), but even for the other two their linear drop-off away from the surface of the cloak corresponds to the linear section of a sine curve.
\subsection{Time-dependence of the Temperature Difference}
Consider the heat equation for some arbitrary domain
\begin{subequations}
\begin{eqnarray}
 \rho C\partial_{t}T	&=& \nabla\cdot(\kappa\nabla T)	\\
		T(\vec{r},0)	&=&	 T_{i}	\\
	  T(\partial r,t)	&=&	 T_{r_{i}}
\end{eqnarray}
\end{subequations}
where $\partial r$ are the boundaries of the domain and the boundary conditions are stationary.
In this case, there exists a steady state profile $\partial_t T^{(SS)}=0$ that uniquely satisfies the boundary conditions.
By linearity, $T=T^{(SS)}+T^{(tr)}$ where
\begin{subequations}
\begin{eqnarray}
 \rho C\partial_{t}T^{(tr)}	&=& \nabla\cdot\left(\kappa\nabla T^{(tr)}\right)	\\
		T^{(tr)}(\vec{r},0)	&=&	 T_{i}-T^{(SS)}	\\
	  T^{(tr)}(\partial r,t)	&=&	 0.
\end{eqnarray}
\end{subequations}
Assuming that the materials are everywhere homogeneous for some coordinate system we can apply a spatial Fourier transform ($\nabla^2 T^{(tr)}\equiv -k^2 T^{(tr)}$) and therefore
\begin{subequations}
\begin{eqnarray}
T^{(tr)}(\vec{r},t)&=&\int T^{(tr)}(\vec{k},0)e^{-k^{2}Dt}e^{-ik\cdot r} \frac{d^{n}k}{(2\pi)^{n/2}} \\
T^{(tr)}(\vec{k},0)&=&\int e^{ik\cdot r}\left[T_{i}-T^{(SS)}(\vec{r})\right] \frac{d^{n}r}{(2\pi)^{n/2}}
\end{eqnarray}
\end{subequations}
where $D=\kappa_0/\rho_0 c_{p0}$ is the thermal diffusivity.
For two systems that differ only in $\rho c_p$ the difference between the two
\begin{equation}
\delta T(\vec{r},t;\Delta D) = \int (\vec{k},0)\left(e^{-k^{2}D_{a}t}-e^{-k^{2}D_{b}t}\right)e^{-ik\cdot r}\frac{T^{(tr)}d^{n}k}{(2\pi)^{n/2}}.
\label{eq:T_tr}
\end{equation}
Note the time dependence is a sum of the difference of exponentials.
This implies that for short times $\delta T$ is approximately linear while for long times it decays exponentially.
In the case that only a single Fourier mode is excited $\delta T$ is separable.
This is also approximately true if a small number of well separated Fourier modes dominate $T^{(tr)}(\vec{k},0)$.

\subsection{Sensitivity of a Cloak to the Inner Boundary}

Following \cite{PC1} we consider a PC that has lost a section of the inner boundary of thickness $\delta$.
Defining the domains $I, II, III$ to be external to the cloak, the cloak, and the interior the boundary conditions (continuity of $T$ and $\hat{n}\cdot\kappa\nabla T$) are
\begin{widetext}
\begin{subequations}
\begin{eqnarray}
a_{l}^{(I)}I_{l}(\sqrt{i}k_{B}b)+b_{l}^{(I)}K_{l}(\sqrt{i}k_{B}b)			&=&
a_{l}^{(II)}I_{l}(\sqrt{i}k_{C}[b-a])+b_{l}^{(II)}K_{l}(\sqrt{i}k_{C}[b-a])	\\
\kappa_{0}k_{B}\left[a_{l}^{(I)}I_{l}^{\prime}(\sqrt{i}k_{B}b)+b_{l}^{(I)}K_{l}^{\prime}(\sqrt{i}k_{B}b)\right]	&=&
\kappa_{r}k_{C}\left[a_{l}^{(II)}I_{l}^{\prime}(\sqrt{i}k_{C}[b-a])+b_{l}^{(II)}K_{l}^{\prime}(\sqrt{i}k_{C}[b-a])\right]	\\
a_{l}^{(III)}I_{l}(\sqrt{i}k_{B}[a+\delta])		&=&
a_{l}^{(II)}I_{l}(\sqrt{i}k_{C}\delta)+b_{l}^{(II)}K_{l}(\sqrt{i}k_{C}\delta)	\\
\kappa_0k_{B}a_{l}^{(III)}I_{l}^{\prime}(\sqrt{i}k_{B}[a+\delta])	&=&
\kappa_{r}k_{C}\left[a_{l}^{(II)}I_{l}^{\prime}(\sqrt{i}k_{C}\delta)+b_{l}^{(II)}K_{l}^{\prime}(\sqrt{i}k_{C}\delta)\right]
\end{eqnarray}
\end{subequations}
\end{widetext}
where $k_B=\sqrt{\omega \rho_0 c_{p0}/\kappa_0}$ and $(b-a)k_C = b k_B$, $a_l^{(I)}$ is the incident field component, $b_l^{(I)}$ is the scattered component, $a_l^{(III)}$ is the penetrating field, and we have expanded our solution using the eigenfunctions found in Sec. 1 ($b_l^{(III)}$ is tautologically 0 since $K_l(0)$ diverges).
Using these definitions of $k$ and $\kappa$ the first conditions become
\begin{widetext}
\begin{subequations}
\begin{eqnarray}
a_{l}^{(I)}I_{l}(\sqrt{i}k_{B}b)+b_{l}^{(I)}K_{l}(\sqrt{i}k_{B}b)&=&
a_{l}^{(II)}I_{l}(\sqrt{i}k_{B}b)+b_{l}^{(II)}K_{l}(\sqrt{i}k_{B}b)\\
a_{l}^{(I)}I_{l}^{\prime}(\sqrt{i}k_{B}b)+b_{l}^{(I)}K_{l}^{\prime}(\sqrt{i}k_{B}b)&=&
a_{l}^{(II)}I_{l}^{\prime}(\sqrt{i}k_{B}b)+b_{l}^{(II)}K_{l}^{\prime}(\sqrt{i}k_{B}b),
\end{eqnarray}
\end{subequations}
\end{widetext}
which, given an arbitrary $b$ implies that $a_l^{(II)}=a_l^{(I)}$ and $b_l^{(II)}=b_l^{(I)}$.
Using the last two boundary conditions and the fact that the Wronskian $\mathcal{W}[I_l(z),K_l(z)]=-1/z$ \cite{AS} gives
\begin{widetext}
\begin{subequations}
\begin{eqnarray}
a_{l}^{(III)	}&=&	\frac{-(\sqrt{i}k_{B}a)^{-1}}{\frac{\delta}{a+\delta}\frac{b}{b-a}K_{l}^{\prime}(\sqrt{i}k_{C}\delta)I_{l}(\sqrt{i}k_{B}[a+\delta])-I_{l}^{\prime}(\sqrt{i}k_{B}[a+\delta])K_{l}(\sqrt{i}k_{C}\delta)}a_{l}^{(I)}\\
b_{l}^{(I)}		&=&		\frac{I_{l}(\sqrt{i}k_{C}\delta)I_{l}^{\prime}(\sqrt{i}k_{B}[a+\delta])-\frac{\delta}{a+\delta}\frac{b}{b-a}I_{l}^{\prime}(\sqrt{i}k_{C}\delta)I_{l}(\sqrt{i}k_{B}[a+\delta])}{\frac{\delta}{a+\delta}\frac{b}{b-a}K_{l}^{\prime}(\sqrt{i}k_{C}\delta)I_{l}(\sqrt{i}k_{B}[a+\delta])-I_{l}^{\prime}(\sqrt{i}k_{B}[a+\delta])K_{l}(\sqrt{i}k_{C}\delta)}a_{l}^{(I)}
\end{eqnarray}
\end{subequations}
\end{widetext}
which can be expanded in the limit $\delta\to0$.
For $l\ne0$ this gives
\begin{widetext}
\begin{subequations}
\begin{eqnarray}
a_{l}^{(III)}	&\approx&	\frac{(\frac{1}{2}\sqrt{i}k_{C}\delta)^{l}}{(l-1)!} \frac{(\sqrt{i}k_{B}a)^{-1}}{l(\frac{1}{2}\sqrt{i}k_{B}a)^{-1}I_{l}(\sqrt{i}k_{B}a)+\frac{1}{2}I_{l}^{\prime}(\sqrt{i}k_{B}a)}a_{l}^{(I)} \\
b_{l}^{(I)}		&\approx&	\frac{2(\frac{1}{2}\sqrt{i}k_{C}\delta)^{2l}}{l[(l-1)!]^{2}}\frac{l(\sqrt{i}k_{B}a)^{-1}I_{l}(\sqrt{i}k_{B}a)-I_{l}^{\prime}(\sqrt{i}k_{B}a)}{4l(\sqrt{i}k_{B}a)^{-1}I_{l}(\sqrt{i}k_{B}a)+I_{l}^{\prime}(\sqrt{i}k_{B}a)}a_{l}^{(I)}.
\end{eqnarray}
\end{subequations}
\end{widetext}
which vanish at $\delta=0$
For $l=0$ this gives
\begin{subequations}
\begin{eqnarray}
a_{0}^{(III)}	&\approx&	-\frac{1}{(\sqrt{i}k_{B}a)I_{0}^{\prime}(\sqrt{i}k_{B}a)\ln k_{C}\delta}a_{0}^{(I)}	\\
b_{0}^{(I)}	&\approx&	\frac{1}{\ln k_{C}\delta}a_{0}^{(I)}
\end{eqnarray}
\end{subequations}
which also vanishes at $\delta=0$ but converges more slowly than the previous case.
For $\omega =0$ repeating the same procedure gives 
\begin{subequations}
\begin{eqnarray}
A_{l}^{(II)}	&=&\left(\frac{b}{b-a}\right)^{l}A_{l}^{(I)}	\\
A_{l}^{(III)}	&=&\frac{2b-2a}{2b-a}\left(\frac{b}{b-a}\frac{\delta}{a}\right)^{l}A_{l}^{(I)}	\\
B_{l}^{(I)}		&=&\frac{-a}{2b-a}\left(\frac{b}{b-a}\delta\right)^{2l}A_{l}^{(I)}\\
B_{l}^{(II)}	&=&\frac{-a}{2b-a}\delta^{2l}A_{l}^{(I)}
\end{eqnarray}
\end{subequations}
for $l\ne0$ and $A_0^{(I)}=A_0^{(II)}=A_0^{(III)}$, $B_0^{(I)}=B_0^{(II)}=0$ for $l=0$.
Thus for a PC ($\delta\to0$) the temperature inside is a constant and the scattering field vanishes.
This confirms that a PC is truly perfect, as expected.

\subsection{Simulations and Experimental Study of the BC}

\begin{figure}[h] \begin{center}
\includegraphics[scale=0.12]{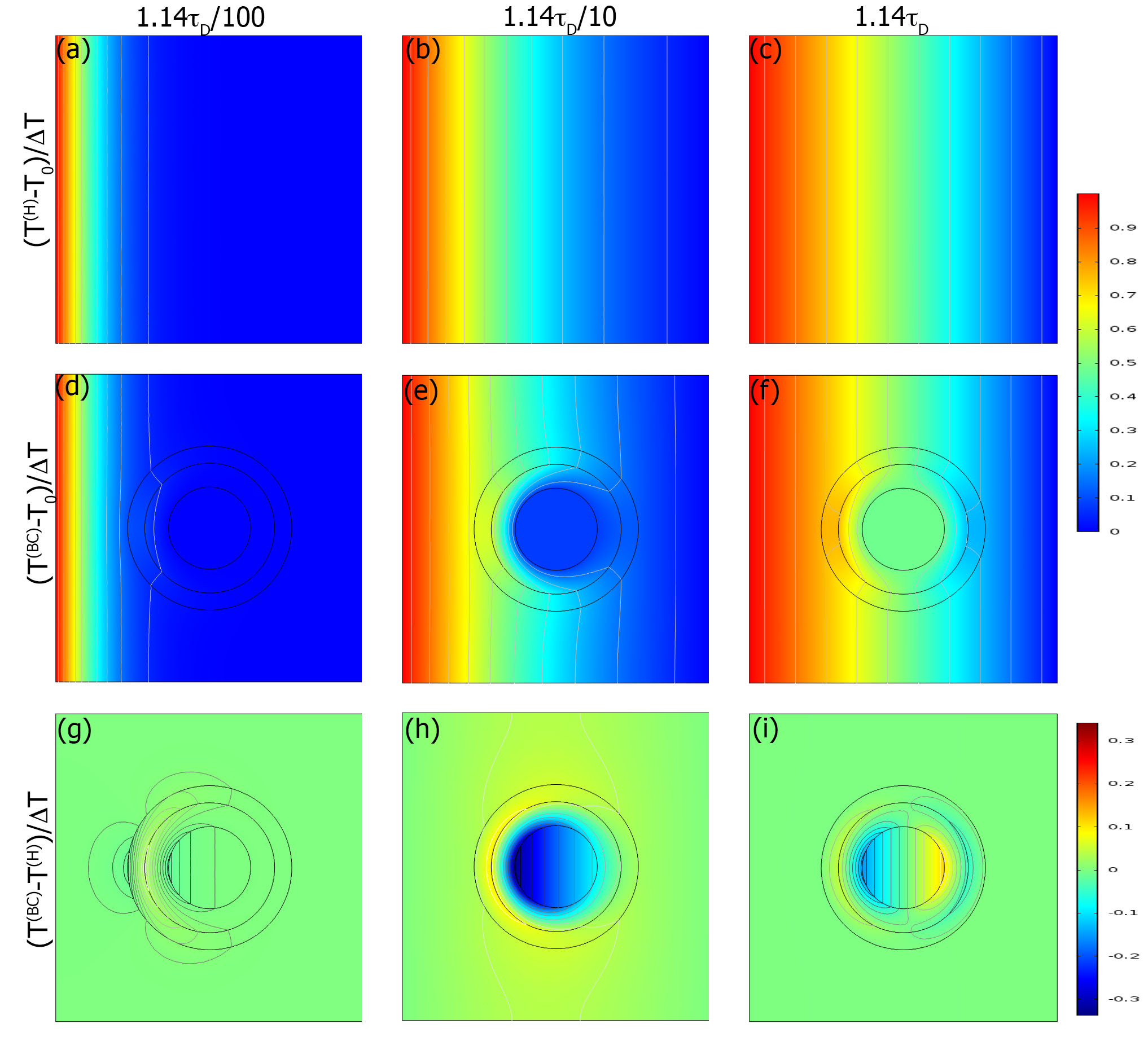}
\caption{\label{fig:BC} Simulated temperature snapshots. Rows correspond to 30s ($1.14\tau_D/100$), 300s ($1.14\tau_D/10$), and 3000s ($1.14\tau_D$) respectively. Columns correspond to the homogeneous case (no cloak), BC, and $T^{(BC)}-T^{(H)}$. Black circles denote the location of the cloak (for reference in the homogeneous case), colored domains are isotherms, and grey lines are constant separation isotherms.
}
\end{center}
\end{figure}

We follow \cite{BC} to model the BC as rectangular domain of dimensions $L=$45 mm by $L_\perp=$45 mm centered around a cloak with hidden region of size $a=$6 mm, first layer of $r_2=$9.5 mm, and second layer of $b=$12 mm.
The background medium is $\kappa_0=2.3 \mathrm{W/m\cdot K}$, $\rho_0 = 2000 \mathrm{kg/m^3}$, and $c_{p0} = 1500\mathrm{J/kg\cdot K}$, the outer layer's medium is $\kappa_1=9.8 \mathrm{W/m\cdot K}$, $\rho_1 = 8440 \mathrm{kg/m^3}$, and $c_{p1} = 400\mathrm{J/kg\cdot K}$, the inner layer's medium is $\kappa_2=0.03 \mathrm{W/m\cdot K}$, $\rho_2 = 50 \mathrm{kg/m^3}$, and $c_{p2} = 1300\mathrm{J/kg\cdot K}$, and the interior medium is $\kappa_3=205 \mathrm{W/m\cdot K}$, $\rho_3 = 2700 \mathrm{kg/m^3}$, and $c_{p3} = 900\mathrm{J/kg\cdot K}$.
This gives a diffusivity of $D_0=\kappa_0/\rho_0c_{p0}= 7.67\cdot10^{-7}\mathrm{m^2/s}$ and diffusion timescale $\tau_{D_0}=L^2/D=2641.3$s.
The initial temperature was 273.15K with thermal baths at 333.15K, and $T_0=$273.15K giving a $\Delta T$ of 60K.
For plotting we use the natural units of $x/L, y/L, t/\tau_{D_0},(T-T_0)/\Delta T$.
The results are shown in Fig. \ref{fig:BC}, confirming that the cloak is visible in the transient response

\begin{figure}[h] \begin{center}
\includegraphics[scale=.4]{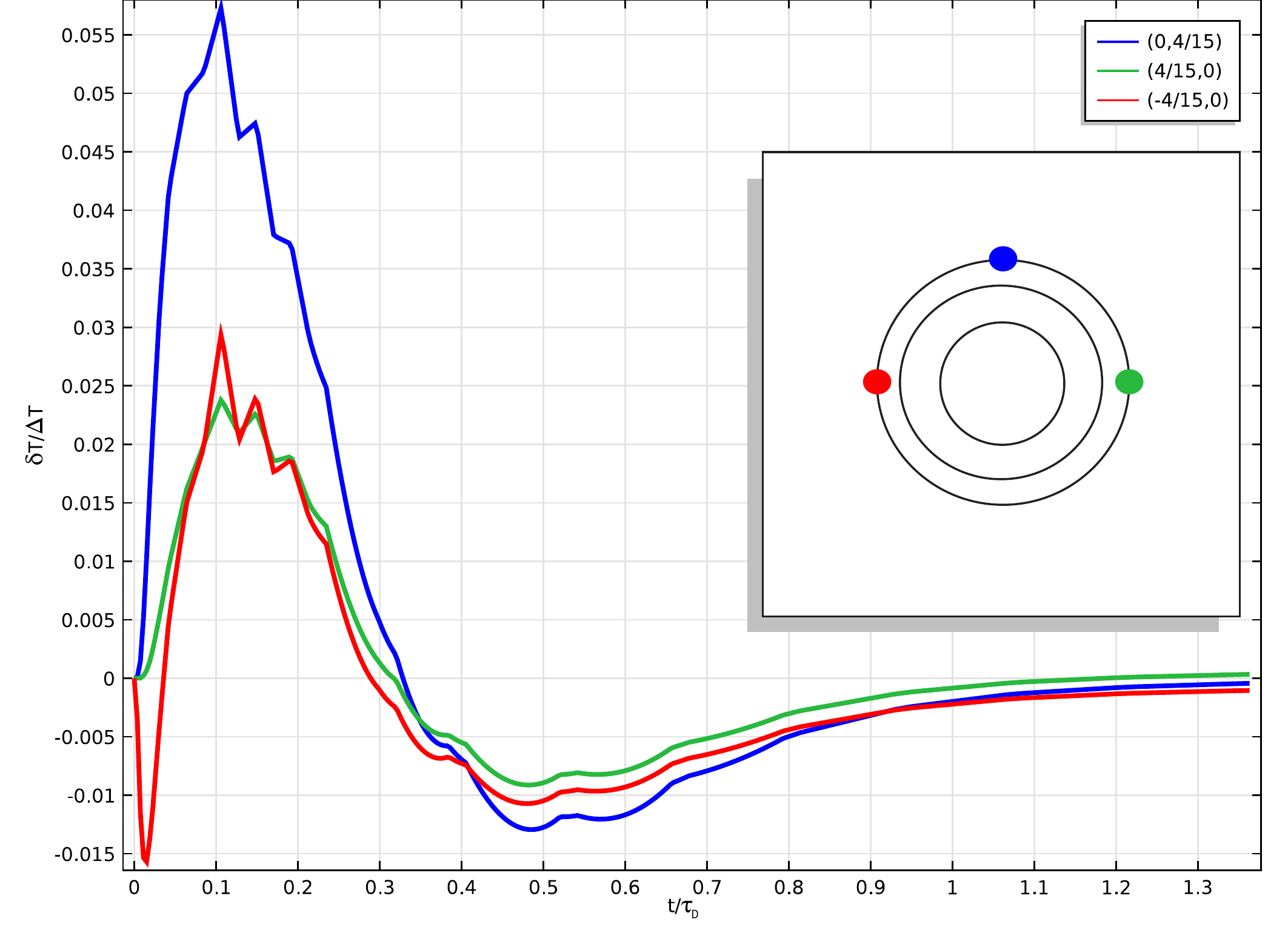}
\caption{\label{fig:BCtime} Temperature deviation $\delta T/\Delta T$ for representative points outside the cloak as a function of time. Color corresponds to different points (see inset for key). 
}
\end{center}
\end{figure}

We also test the time-dependence of $\delta T$ using Fig. \ref{fig:BCtime}.
As expected, the time-dependence is a sum of exponential terms like those predicted in eq. \ref{eq:T_tr}.
Because of the additional boundaries in this system we see that there are more Fourier modes excited.
What's more, the addition of these Fourier modes implies that the solution is not fully separable.
This is clear from the separation of $\delta T$ at the nearest point around the cloak to the heat source.
This can also be seen with the initial plot of $\delta T$ in Fig. \ref{fig:BC} where there is initially relative cooling outside the cloak that is not found elsewhere along it's surface. 

\begin{figure}[h] \begin{center}
\includegraphics[scale=0.15]{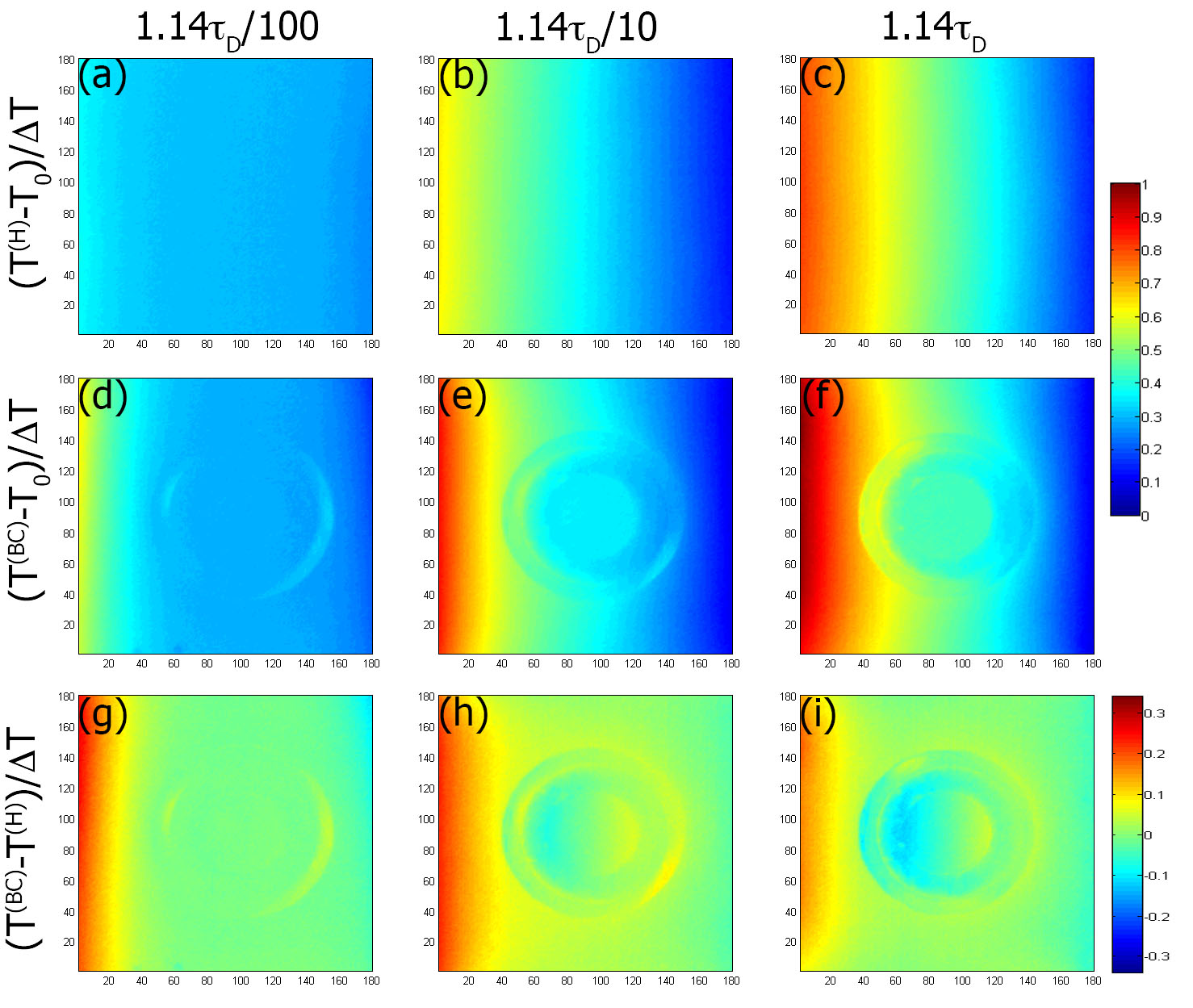}
\caption{\label{fig:BCE} Experimental temperature snapshots. Rows correspond to 30s ($1.14\tau_D/100$), 300s ($1.14\tau_D/10$), and 3000s ($1.14\tau_D$) respectively. Columns correspond to the homogeneous case (no cloak), BC, and $T^{(BC)}-T^{(H)}$. Black circles denote the location of the cloak (for reference in the homogeneous case), colored domains are isotherms, and grey lines are constant separation isotherms.
}
\end{center}
\end{figure}

To verify our simulations we follow the procedure of Fig. \ref{fig:BC} with an experimental realization of the BC and its homogeneous background.
Since the temperatures at each boundary are not perfectly fixed, we normalize the data using the infimum of $T_0=$285.08K and supremum of $T=$325.96K giving $\Delta T=$40.88K.
Using the normalization $(T-T_0)/\Delta T$ shows good agreement with the theoretical result.
Results are plotted in Fig. \ref{fig:BCE}.
There is a slight discrepancy in the temperature deviation between the simulations and experiment.
This is due to a slight difference in temperature gradients applied to the BC and homogeneous cases.
Hence, this effect is strongest at the boundaries of the system and more negligible near the cloak itself.

\subsection{Detecting the Interior of a BC}

\begin{figure}[h] \begin{center}
\includegraphics[scale=.4]{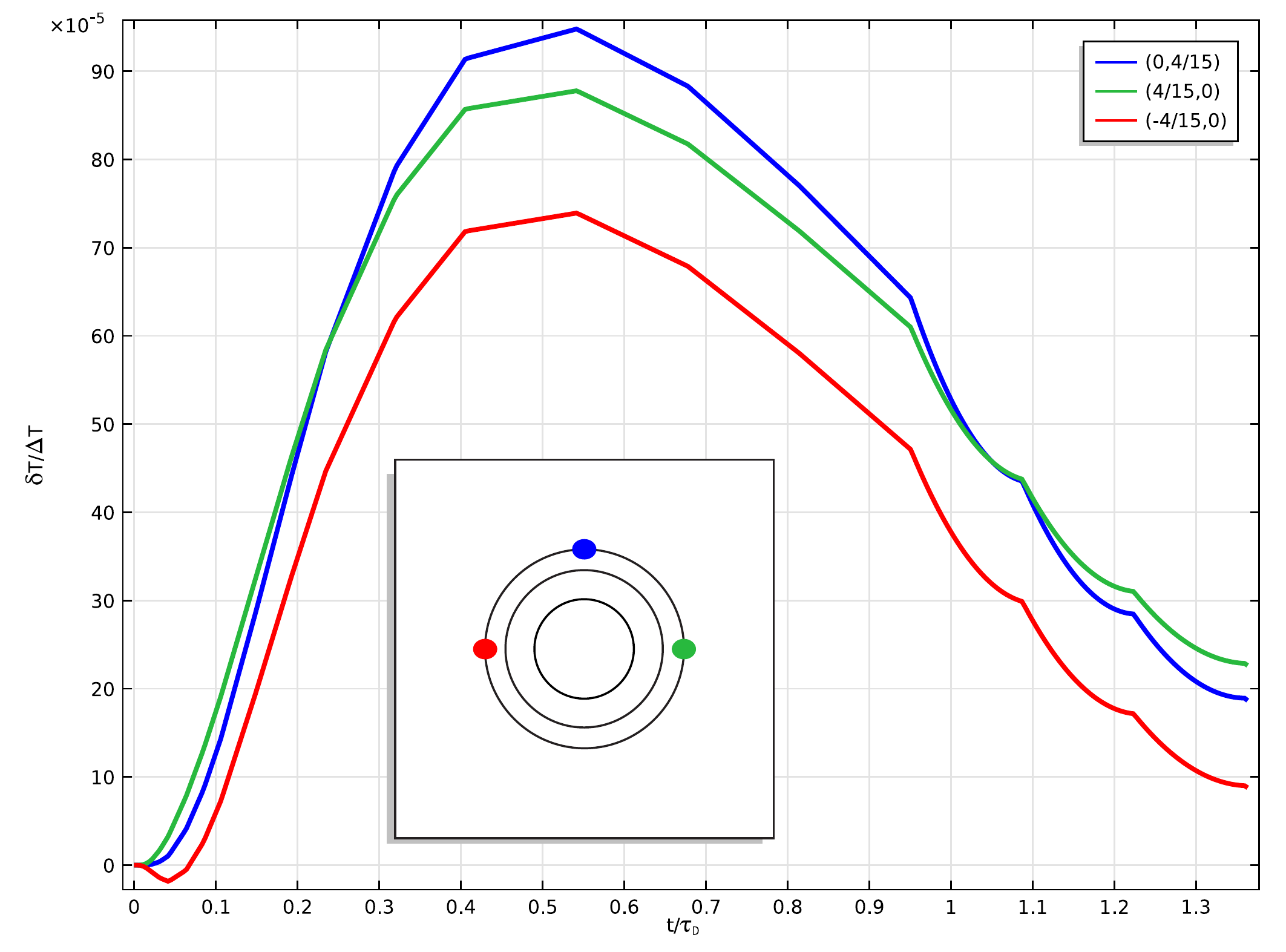}
\caption{\label{fig:BCobj} Temperature deviation $(T^{(cloak+object)}-T^{(cloak)})/\Delta T$ for representative points outside the cloak as a function of time. Color corresponds to different points (see inset for key). 
}
\end{center}
\end{figure}

To compare the effect of different objects hidden within a BC, we repeat the simulation of the previous section with the cloaked object being background.
Subtracting this solution from previous case (with an object) gives the component of the signal that's due to the cloaked object's impedance.
Plotting the time dependence for representative points around the cloak's outer edge as a function of time in Fig. \ref{fig:BCobj} shows that this is an incredibly small signal (less than $10^{-3}$) even at it's peak.
For a thermometer with a sensitivity of 0.2K, detecting this effect would require an applied temperature difference of at least 200K at a minimum (more if the detector is away from the surface of the cloak).
Since this is a larger temperature difference than is typically applied in experiments the interior cloaked in practice (but not in principle).

\begin{figure}[h] \begin{center}
\includegraphics[scale=.375]{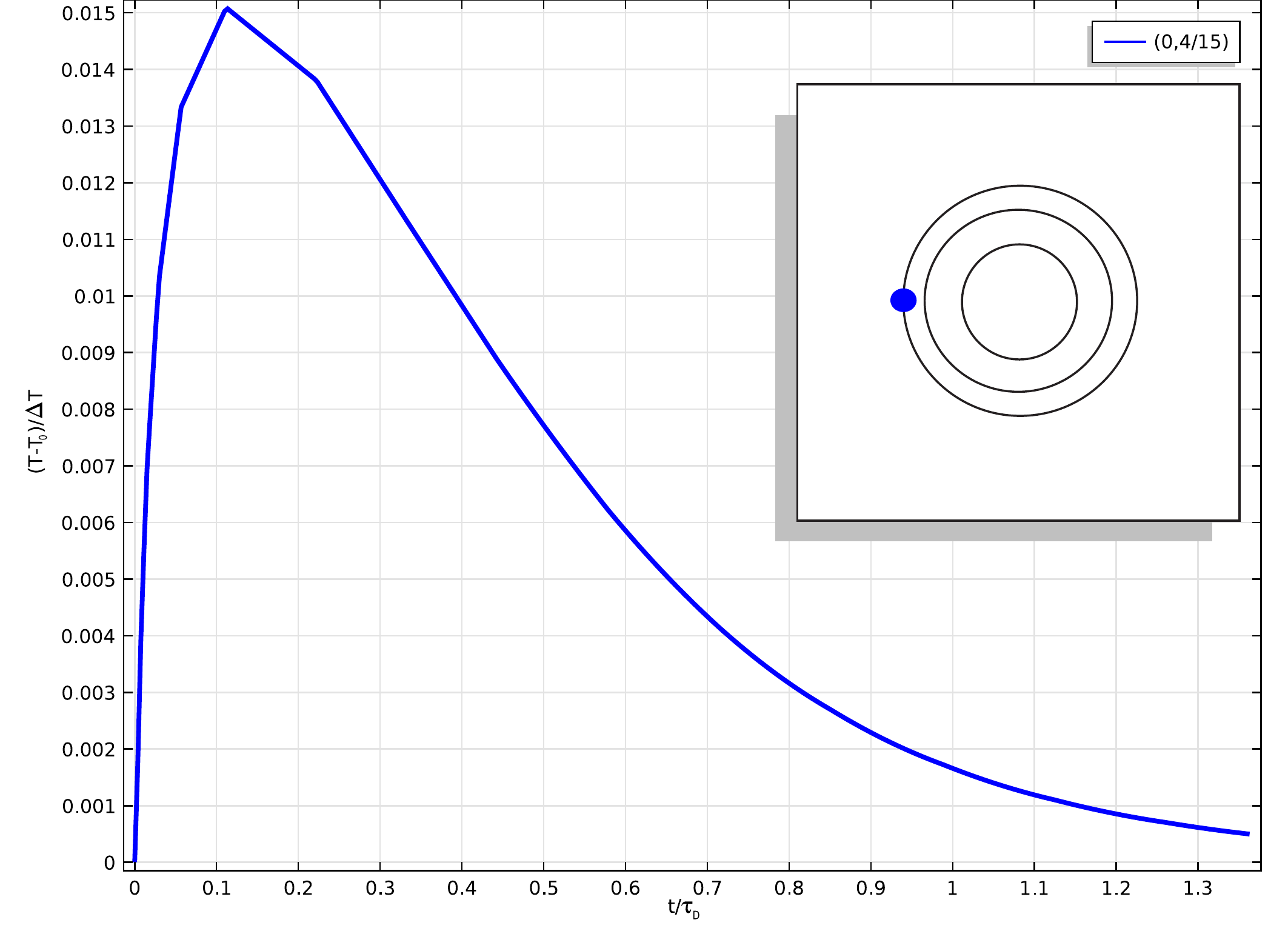}
\includegraphics[scale=.375]{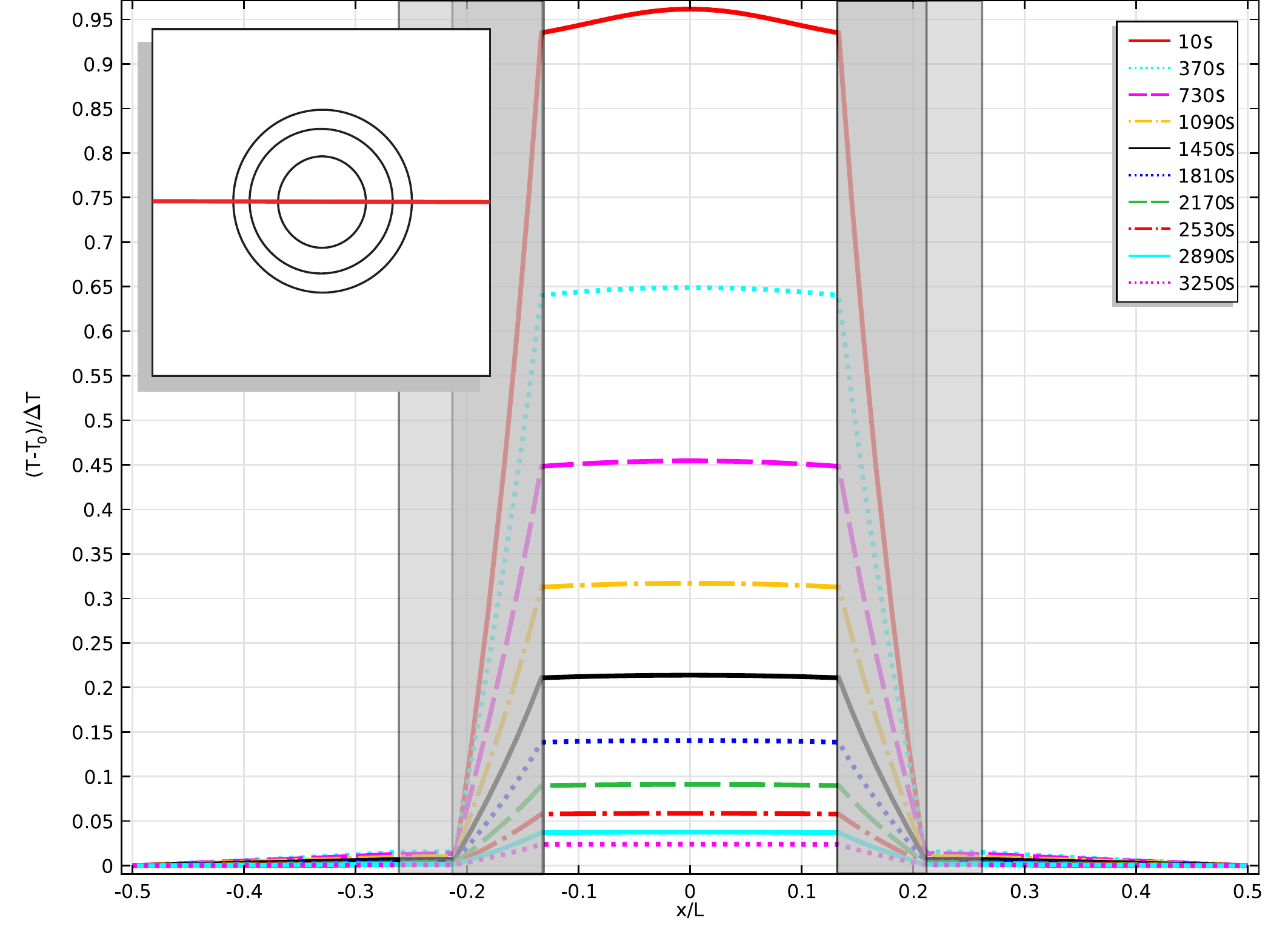}
\caption{\label{fig:BCT0} Temperature response of the BC used as an insulator. Plot (a) shows the time dependence of a representative point outside the cloak while plot (b) shows the spatial dependence of a slice through the cloak (see inset) at various times (in seconds). The fluctuation near the inner boundary is a numerical artifact of the discontinuity in temperature.
}
\end{center}
\end{figure}

As for the ability of a cloak to insulate a cloaked object and thus disguise the temperature profile, it is helpful to use different boundary and initial conditions.
Instead of applying a thermal gradient across the boundaries, the cloaked region is initially set to 60K above the background (and cloak) at 273.15K (These values are then rescaled to 1 and 0).
If the cloak were perfect this initial condition would persist indefinitely.
Instead, we find in Fig. \ref{fig:BCT0} that the temperature inside the cloak decays to the background value.
Because the thermal baths are at fixed temperature and perfectly absorb heat flux, energy is not conserved in this simulation and so the steady state should have all the heat removed from the cloak.
Assuming a similar level of sensitivity of our thermometer, the temperature difference required to observe this is about 14.3K.
This is well within experimental detectability, albeit an order of magnitude smaller than detecting the presence of a cloak (which requires a gradient of at least 3.64K)

%\section{Acknowledgements}
%This material is based upon work supported by the National Science Foundation %Graduate Research Fellowship under Grant No. 1122374. 

%\section{Author contributions}
%S.R.S. proposed the project and performed the analytic calculations and %computational simulations. X.B. performed the experiment. S.R.S and B.L. analyzed %the results and wrote the manuscript. B.Z and X.Z. supervised the project.

%\section{Competing financial interest}
%Authors declare no competing financial interest.

\end{document}